\documentclass[fleqn,a4paper,12pt]{article}
\usepackage[utf8]{inputenc}
\usepackage{amssymb}
\usepackage{amsmath}
\usepackage[dvips]{graphicx}
\usepackage{placeins}
\usepackage{lscape}
\usepackage{a4}
\usepackage{epsfig}

\catcode`@=11 \long
\def\@caption#1[#2]#3{\par\addcontentsline{\csname
ext@#1\endcsname}{#1} {\protect\numberline{\csname
the#1\endcsname}{\ignorespaces #2}} \begingroup \small
\@parboxrestore \@makecaption{\csname fnum@#1\endcsname}
{\ignorespaces #3}\par \endgroup} \catcode`@=12

\renewcommand{\bar}{\overline}

\newcommand{\ddiag}[3]{\mathrm{diag}\left\{#1,#2,#3\right\}}

\newcommand{\p}{\partial}

\newcommand{\CZ}{\mathbb{C}}
\newcommand{\R}{\mathbb{R}}

\DeclareMathOperator{\sech}{sech}
\DeclareMathOperator{\csch}{csch}

\begin{document}
\allowdisplaybreaks
\begin{titlepage} \vskip 2cm

\begin{center}
{\Large\bf Three-dimensional Matrix Superpotentials}
\vskip 3cm {\bf {Yuri Karadzhov\footnote{E-mail: {\tt yuri.karadzhov@gmail.com}}} \vskip 5pt
{\sl Institute of Mathematics, National Academy of Sciences of Ukraine,\\
3 Tereshchenkivs'ka Street, Kyiv-4, Ukraine, 01601\\}}
\end{center}
\vskip .5cm \rm
\begin{abstract}
This article considers the classification of matrix superpotentials that corresponds to exactly
solvable systems of Schr\"odinger equations. Superpotentials of the following form are considered:
$W_k = kQ + P + \frac1kR$, where $k$ --- parameter,
$P, Q$ and $R$ --- hermitian matrices, that depend on a variable $x$.
The list of three-dimensional matrix superpotentials is presented explicitly.
\end{abstract}

\end{titlepage}

\section{Introduction \label{intro}}

Supersymmetric quantum mechanics presents a powerful tool for finding exact solutions
to problems described by Schr\"odinger equations \cite{Witten}.
Discrete symmetry between a Hamiltonian and its superpartner, called shape-invariance \cite{Gen}
helps solve the spectral problem by means of algebraic methods.

Unfortunately, the class of known problems that satisfy the shape-invariance condition is rather limited.
However, it almost entirely covers the cases when the corresponding Schr\"odinger equation can be exactly
integrable and has an explicitly presented potential.

A well known paper by Cooper et al \cite{Khare} presents a classification of scalar potentials
that correspond to exactly solvable Schr\"odinger equations. Matrix potentials arise in many
physics problems. For instance, the motion of neutral nonrelativistic fermion that interacts
anomaly with the magnetic field generated by a thin current-carrying wire, is described by a model
with a matrix potential. This model was proposed by Pron'ko and Stroganov \cite{Pron}.
Other good examples of problems with matrix potentials can be found in works \cite{pl1}-\cite{pl3}
that describe crystal structures in the Gross-Neve model. Some two-dimensional matrix potentials
can be found in \cite{tk1}, \cite{tk2}. Particular cases of matrix potentials can be also found
in works by \cite{Andr1}-\cite{Rodr}.
In \cite{Fu} a particular class of two-dimensional matrix potentials is presented.

A systematic study of the problem was introduced in \cite{kar1}, and \cite{kar2},
where presented a classification of superpotentials of the form $W_k = kQ + P + \frac1kR$,
where $k$ --- parameter, $P, Q$ and $R$ --- Hermitian matrices and one of the following conditions is held:
$Q$ is proportional to a unit matrix, or $R$ is equal to zero. In \cite{kar3} a wide class
of matrix superpotentials was considered, however a full classification was not made.

This paper considers three-dimensional superpotentials and presents their classification.

\section{Solving the spectral problem \label{spectrprob}}

Let us consider the following spectral problem
\begin{gather}
\label{specprob}
H_k\psi = E_k\psi,
\end{gather}
where $H_k$ --- Hamiltonian with a matrix potential, $E_k$ and $\psi$ --- its eigenvalues
and eigenfunctions correspondingly.
In the Schr\"odinger equation Hamiltonian has the following form
\begin{gather}
\label{schrodham}
H_k = -\frac{\p^2}{\p x^2} + V_k(x),
\end{gather}
where $V_k(x)$ --- matrix potential dependent on the parameter $k$ and the variable $x$.

Let us assume that Hamiltonian can be factorized as follows
\begin{gather}
\label{hfact}
H_k = a_k^\dag a_k + c_k,
\end{gather}
where $c_k$ --- scalar function of $k$, that vanishes with a corresponding member in the Hamiltonian.
From here on, we will drop the sign of unit matrix $I$ and write $c_k$ instead of $c_kI$.
As \cite{kar2} mentions it is enough to consider the operators $a_k$ and $a_k^\dag$ of the form
\begin{gather}
\label{standrep}
a_k=\frac{\p}{\p x}+W_k(x),\quad a_k^\dag =-\frac{\p}{\p x}+W_k(x),
\end{gather}
where $W_k$ --- hermitian matrix, that is called a superpotential.

As $W_k$ is hermitian, operators $a_k$ and $a_k^\dag$ are hermitian-conjugate,
which allows to find the ground state of the spectral problem (\ref{specprob}),
by simply solving the first order differential equation.
Indeed, multiplying the expression
\begin{gather}
\label{grsteqnonsimp}
a_k^\dag a_k\psi = 0
\end{gather}
on the left by the hermitian-conjugate spinor $\psi^\dag$ and integrating
the expression on the real line~$\R$ we get
\begin{gather}
\label{grsteqnorm}
||a_k\psi||_2 = 0,
\end{gather}
where $||\cdot||_2$ denotes the norm in $L_2(\R)$. Hence,
\begin{gather}
\label{grsteq}
a_k\psi = 0.
\end{gather}
The square-integrable function $\psi_k^0(x)$, that is a normalized solution to the equation~(\ref{grsteq}),
is an eigenfunction of the Hamiltonian, that corresponds to the eigenvalue $E_k^0 = c_k$,
and is called a ground state of the system (\ref{specprob}).

Suppose the system satisfies the shape-invariance condition:
\begin{gather}
\label{hshapeinv}
H_k^+ = H_{k+1},
\end{gather}
where $H_k^+$ is a Hamiltonian's superpartner, that is defined by the following formula
\begin{gather}
\label{superpartner}
H_k^+ = a_ka_k^\dag + c_k.
\end{gather}
This condition allows to fully discover the spectrum by a series of algebraic operations,
knowing the eigenvalue of the system $\psi_k^0(x)$. Indeed, if we use the condition  (\ref{hshapeinv}),
it is easy to show that the function
\begin{gather}
\label{exstone}
\psi_k^1(x) = \frac{a_k^\dag\psi_{k+1}^0(x)}{||a_k^\dag\psi_{k+1}^0(x)||_2}
\end{gather}
is a Hamiltonian's eigenfunction with an eigenvalue $E_k^1 = c_{k+1}$. It is called
the first exited state of the system (\ref{specprob}). Analogously, by induction we prove that  the function
\begin{gather}
\label{exstn}
\psi_k^n(x)=\frac{a_k^\dag a_{k+1}^\dag\cdots a_{k+n-1}^\dag\psi_{k+n}^0(x)}{||a_k^\dag a_{k+1}^\dag\cdots a_{k+n-1}^\dag\psi_{k+n}^0(x)||_2}
\end{gather}
is a Hamiltonian's eigenfunction with an eigenvalue $E_k^n = c_{k+n}$. It is called the
$n^{th}$ exited state  of the system (\ref{specprob}).

Consequently, if the system of Schr\"odinger equations (\ref{specprob}) satisfies
the shape-invariance condition, it can be exactly integrated.

\section{The classification problem \label{deteq}}

Since the shape-invariant potentials correspond to the exactly integrable systems of Schr\"odinger equations,
it would be interesting to widen the class. Let us find all the Hamiltonians that allow factorization
(\ref{hfact}) and satisfy the shape-invariance condition~(\ref{hshapeinv}). In terms of the superpotential
these conditions can be written through a single equation
\begin{gather}
\label{shapeinv}
W_k^2 + W'_k = W_{k+1}^2 - W'_{k+1} + C_k,
\end{gather}
where $C_k = c_{k+1} - c_k$. Thus, to solve the given problem, it is enough to find all the superpotentials,
that satisfy the equation (\ref{shapeinv}).

In the general form, the problem is rather complex. However, it can be solved if we consider
superpotentials of a chosen class. This paper considers superpotentials of the form
\begin{gather}
\label{superpot}
W_k = kQ + P + \frac1kR,
\end{gather}
where $P, Q$ and $R$ --- Hermitian matrices of $3\times3$ dimension.
Additionally, let us consider that $Q$ is not proportional to a unit matrix and $R$ is not a zero matrix.
These cases were considered in \cite{kar1}, \cite{kar2}.

We are interested in the irreducible superpotentials, i.e. the ones that can not be reduced to
a block-diagonal form by means of an unitary transformation, that does not depend on variable $x$.
In the reducible superpotentials case, the problem is divided into a set of similar problems of a smaller dimension.

In the following section the equation describing unknown matrices $P, Q$ and $R$ is obtained and solved,
given the corresponding superpotentials satisfy the equation~(\ref{shapeinv}).

\section{The determining equations \label{deteqsol}}

To get a system of determining equations, let us substitute the expression (\ref{superpot}),
for the superpotential, into the equation (\ref{shapeinv}) and separate the variables. Hence,
\begin{gather}
\label{classeq}
Q' = Q^2 + \nu, \\
\label{classep}
P' = \frac12\{P,Q\} + \mu, \\
\label{classer1}
R' = 0, \\
\label{classer2}
R^2 = \omega^2, \\
\label{classepr}
\{P,R\} + \varkappa = 0, \\
\label{classec}
C_k = 2\mu +(2k+1)\nu - \frac\varkappa{k(k+1)} + \frac{(2k+1)\omega^2}{k^2(k+1)^2},
\end{gather}
where $\nu, \mu, \omega, \varkappa$ --- arbitrary real constants.

As previously shown in \cite{kar2}, matrix $Q$ can be diagonalized by means of a unitary transformation
that does not depend on $x$. Then the expression (\ref{classeq}) is reduced to a system
\begin{equation}
\label{classeqsys}
q_i' = q_i^2 + \nu,\quad i = 1\dots3,
\end{equation}
where $q_i$ --- diagonal elements of the matrix $Q$.

It has the following solutions:
\begin{gather}
\label{solqi}
\begin{split} &
 q_i=\lambda\tan(\lambda x+\gamma_i),\quad i=1\dots3,\quad \nu=\lambda^2; \\ &
 q_i=\left[\begin{array}{ll}
            -\lambda\tanh(\lambda x+\gamma_i),& i=1\dots m \\[1 ex]
        -\lambda\coth(\lambda x+\gamma_i),& i=m+1\dots l \\[1 ex]
        \pm\lambda,& i=l+1\dots3
           \end{array}\right.,\quad \nu=-\lambda^2; \\ &
 q_i=\left[\begin{array}{ll}
            -\frac1{x+\gamma_i},& i=1\dots m \\[1 ex]
        0,& i=m+1\dots3
           \end{array}\right.,\quad \nu=0,
\end{split}
\end{gather}
where $\gamma_i\in\R, i = 1\dots3$ --- integral constants, $m, l = 0\dots4$,
gang $a\dots b$, where $a > b$ is considered empty.

As matrix $Q$ is diagonal, then the linear equation (\ref{classep})
is dissolved and can be solved element-wise:
\begin{itemize}
\item If $\nu=\lambda^2$
\end{itemize}
\begin{gather}
\label{solP1}
\begin{split} &
 p_{ii}=\frac{\mu}{\lambda}\tan(\lambda x+\gamma_i)+\varphi_{ii}\sec(\lambda x+\gamma_i),\quad i=1\dots3 \\ &
 p_{ij}=\varphi_{ij}\sqrt{\sec(\lambda x+\gamma_i)\sec(\lambda x+\gamma_j)},\quad i=1\dots3, j=1\dots3
\end{split}
\end{gather}
\begin{itemize}
\item If $\nu=-\lambda^2$
\end{itemize}
\begin{gather}
\label{solP2}
\begin{split} &
 p_{ii}=\left[\begin{array}{ll}
  -\frac{\mu}{\lambda}\tanh(\lambda x+\gamma_i)+\varphi_{ii}\sech(\lambda x+\gamma_i),& i=1\dots m \\[1 ex]
  -\frac{\mu}{\lambda}\coth(\lambda x+\gamma_i)+\varphi_{ii}\csch(\lambda x+\gamma_i),& i=m+1\dots l \\[1 ex]
  \pm\frac{\mu}{\lambda}+\varphi_{ii}\exp(\pm\lambda x),& i=l+1\dots3
 \end{array}\right. \\ &
 p_{ij}=\left[\begin{array}{ll}
  \varphi_{ij}\sqrt{\sech(\lambda x+\gamma_i)\sech(\lambda x+\gamma_j)},& i=1\dots m, j=1\dots m \\[1 ex]
  \varphi_{ij}\sqrt{\sech(\lambda x+\gamma_i)\csch(\lambda x+\gamma_j)},& i=1\dots m, j=m+1\dots l \\[1 ex]
  \varphi_{ij}\sqrt{\sech(\lambda x+\gamma_i)\exp(\pm\lambda x)},& i=1\dots m, j=l+1\dots3 \\[1 ex]
  \varphi_{ij}\sqrt{\csch(\lambda x+\gamma_i)\csch(\lambda x+\gamma_j)},& i=m+1\dots l, j=m+1\dots l \\[1 ex]
  \varphi_{ij}\sqrt{\csch(\lambda x+\gamma_i)\exp(\pm\lambda x)},& i=m+1\dots l, j=l+1\dots3 \\[1 ex]
  \varphi_{ij}\exp(\pm\lambda x), \quad(\star)& i=l+1\dots3, j=l+1\dots3 \\[1 ex]
  \varphi_{ij}, \quad(*)& i=l+1\dots3, j=l+1\dots3
 \end{array}\right.
\end{split}
\end{gather}
Let us use the $(\star)$ formula, when the corresponding diagonal elements $q_i$ and $q_j$
have same signs, and $(*)$, if the signs are different.
\begin{itemize}
\item If $\nu=0$
\end{itemize}
\begin{gather}
\label{solP3}
\begin{split} &
 p_{ii}=\left[\begin{array}{ll}
              \dfrac{\varphi_{ii}-\dfrac{\mu x}2(x+2\gamma_i)}{x+\gamma_i},& i=1\dots m \\[1 ex]
          -\mu x+\varphi_{ii},& i=m+1\dots3
              \end{array}\right. \\ &
 p_{ij}=\left[\begin{array}{ll}
              \dfrac{\varphi_{ij}}{\sqrt{(x+\gamma_i)(x+\gamma_j)}},& i=1\dots m, j=1\dots m \\[1 ex]
          \dfrac{\varphi_{ij}}{\sqrt{x+\gamma_i}},& i=1\dots m, j=m+1\dots3 \\[1 ex]
          \varphi_{ij},& i=m+1\dots3, j=m+1\dots3
              \end{array}\right.
\end{split}
\end{gather}
where $\varphi_{ji}=\bar{\varphi_{ij}}\in\CZ$ --- integral constants, and $m$ and $l$ in the gangs
correspond to such in the formula (\ref{solqi}).

Through equations (\ref{classer1}), and (\ref{classer2}) we can conclude that $R=(r_{ij})$ is a constant matrix,
square of which is proportional to a unit matrix.

The last equation (\ref{classepr}) imposes extra conditions onto the constants $\mu, \varkappa$ and
integral constants $\varphi_{ij}, r_{ij}$. Let us show that, in the cases when $Q$ is not a constant matrix,
the following condition has to be held
\begin{equation}
\label{coeffcond}
\mu = 0, \quad \varkappa = 0.
\end{equation}
Let us consider the elements $\{P,R\}_{ij}$ in the equation (\ref{classepr}),
that correspond to $q_i$ --- a nonconstant element of the matrix $Q$:
\begin{gather}
\label{equmukappa}
\begin{split} &
4\mu r_{ii}\xi_i(x)+\sum_{p=1}^3(r_{ip}\bar{\varphi_{ip}}+\bar{r_{ip}}\varphi_{ip})\eta_{ip}(x)=-\varkappa, \\ &
2\mu\bar{r_{ij}}(\xi_i(x)+\xi_j(x))+\sum_{p=1}^3(\bar{\varphi_{ip}}r_{jp}\eta_{ip}(x)+\varphi_{jp}\bar{r_{ip}}\eta_{jp}(x))=0, \quad j=1\dots3,
\end{split}
\end{gather}
where $\eta_{ij}$ --- multipliers of $\varphi_{ij}$ in the matrix $P$; $\xi_i,\ i=1\dots3$ defined by the formula
\begin{gather}
\xi_i(x)=\left[\begin{array}{ll}
\frac1{\lambda}\tan(\lambda x+\gamma_i),& \nu=\lambda^2 \\[1 ex]
-\frac1{\lambda}\tanh(\lambda x+\gamma_i),& \nu=-\lambda^2 \\[1 ex]
-\frac1{\lambda}\coth(\lambda x+\gamma_i),& \nu=-\lambda^2 \\[1 ex]
-\dfrac{x(x+2\gamma_i)}{2(x+\gamma_i)},& \nu=0
\end{array}\right.,
\end{gather}
and $\xi_j$ similar to $\xi_i$, or the constant.

Since $\xi_i(x), \eta_{ij}(x)$ and 1 are linearly independent with each other, then the system~(\ref{equmukappa})
will be consistent if only $\mu$ and $\varkappa$ are equal to zero, or the entire column~$r_{ij}$ of the matrix $R$
is equal to zero. However, in the last case matrix $R$ is a singular matrix, which contradicts
the condition (\ref{classer2}), hence (\ref{coeffcond}) is proven.

The achieved conditions significantly simplify the equation (\ref{classepr}) and allow to solve it element-wise,
simplifying the results by means of unitary transformations that do not depend on $x$.

In the following section the solutions to the (\ref{classepr}) equation and results (\ref{solqi})---(\ref{coeffcond}) are collected
and presented in the form of a list of the three-dimensional superpotentials.

\section{Three-dimensional matrix superpotentials \label{threedim}}

Let us write down matrices $P, Q$ and $R$ separately for convenience.
For matrices $P$ and $R$ basis of the Gell-Man type is used:
\begin{gather}
\label{threebases}
\begin{split} &
e_1=\begin{pmatrix}
    0 & 1 & 0 \\
    1 & 0 & 0 \\
    0 & 0 & 0
    \end{pmatrix}
e_6=\begin{pmatrix}
    0 & 0 & 0 \\
    0 & 0 & 1 \\
    0 & 1 & 0
    \end{pmatrix}
e_2=\begin{pmatrix}
    0 & -i & 0 \\
    i & 0 & 0 \\
    0 & 0 & 0
    \end{pmatrix}
e_5=\begin{pmatrix}
    0 & 0 & -i \\
    0 & 0 & 0 \\
    i & 0 & 0
    \end{pmatrix} \\[1 ex] &
e_4=\begin{pmatrix}
    0 & 0 & 1 \\
    0 & 0 & 0 \\
    1 & 0 & 0
    \end{pmatrix}
e_8=\begin{pmatrix}
    0 & 0 & 0 \\
    0 & 0 & 0 \\
    0 & 0 & 1
    \end{pmatrix}
e_3=\begin{pmatrix}
    1 & 0 & 0 \\
    0 & -1 & 0 \\
    0 & 0 & 0
    \end{pmatrix}
e_7=\begin{pmatrix}
    0 & 0 & 0 \\
    0 & 0 & -i \\
    0 & i & 0
    \end{pmatrix}
\end{split}
\end{gather}
and $e_0$ defines unit matrix $I$. And for matrix $Q$ we use a diagonal form
$Q=\ddiag{q_1}{q_2}{q_3}=\frac{q_1+q_2}2e_0+\frac{q_1-q_2}2e_3+\frac{2q_3-q_1-q_2}2e_8$.

The list of nonequivalent representations of the matrix $Q$:
\begin{itemize}
\item If $\nu=-\lambda^2$
\end{itemize}
\begin{gather}
\label{threeQ1}
Q=\ddiag{
  \lambda\tan(\lambda x+\gamma_1)}{
  \lambda\tan(\lambda x+\gamma_2)}{
  \lambda\tan(\lambda x+\gamma_3)
  }
\end{gather}
\begin{itemize}
\item If $\nu=\lambda^2$
\end{itemize}
\begin{gather}
\label{threeQ21}
Q=\ddiag{
  -\lambda\tanh(\lambda x+\gamma_1)}{
  -\lambda\tanh(\lambda x+\gamma_2)}{
  -\lambda\tanh(\lambda x+\gamma_3)
  } \\[1 ex]
\label{threeQ22}
Q=\ddiag{
  -\lambda\coth(\lambda x+\gamma_1)}{
  -\lambda\coth(\lambda x+\gamma_2)}{
  -\lambda\coth(\lambda x+\gamma_3)
  } \\[1 ex]
\label{threeQ23}
Q=\ddiag{
  -\lambda\tanh(\lambda x+\gamma_1)}{
  -\lambda\tanh(\lambda x+\gamma_2)}{
  -\lambda\coth(\lambda x+\gamma_3)
  } \\[1 ex]
\label{threeQ24}
Q=\ddiag{
  -\lambda\coth(\lambda x+\gamma_1)}{
  -\lambda\coth(\lambda x+\gamma_2)}{
  -\lambda\tanh(\lambda x+\gamma_3)
  } \\[1 ex]
\label{threeQ25}
Q=\ddiag{
  -\lambda\tanh(\lambda x+\gamma_1)}{
  -\lambda\tanh(\lambda x+\gamma_2)}{
  \pm\lambda
  } \\[1 ex]
\label{threeQ26}
Q=\ddiag{
  -\lambda\coth(\lambda x+\gamma_1)}{
  -\lambda\coth(\lambda x+\gamma_2)}{
  \pm\lambda
  } \\[1 ex]
\label{threeQ27}
Q=\ddiag{
  \pm\lambda}{
  -\lambda\tanh(\lambda x+\gamma_2)}{
  -\lambda\coth(\lambda x+\gamma_3)
  } \\[1 ex]
\label{threeQ28}
Q=\ddiag{
  \pm\lambda}{
  \pm\lambda}{
  -\lambda\tanh(\lambda x+\gamma_3)
  } \\[1 ex]
\label{threeQ29}
Q=\ddiag{
  \pm\lambda}{
  \pm\lambda}{
  -\lambda\coth(\lambda x+\gamma_3)
  } \\[1 ex]
\label{threeQ210}
Q=\ddiag{
  \lambda}{
  -\lambda}{
  -\lambda\tanh(\lambda x+\gamma_3)
  } \\[1 ex]
\label{threeQ211}
Q=\ddiag{
  \lambda}{
  -\lambda}{
  -\lambda\coth(\lambda x+\gamma_3)
  }
\end{gather}
\begin{itemize}
\item If $\nu=0$
\end{itemize}
\begin{gather}
\label{threeQ31}
Q=\ddiag{
  -\frac1{x+\gamma_1}}{
  -\frac1{x+\gamma_2}}{
  -\frac1{x+\gamma_3}
  } \\[1 ex]
\label{threeQ32}
Q=\ddiag{
  -\frac1{x+\gamma_1}}{
  -\frac1{x+\gamma_2}}{
  0
  } \\[1 ex]
\label{threeQ33}
Q=\ddiag{
  0}{
  0}{
  -\frac1{x+\gamma_3}
  }
\end{gather}
Where two cases are considered: $\gamma_1, \gamma_2, \gamma_3$ --- all are different and $\gamma_1=\gamma_2\neq \gamma_3$.

The list of nonequivalent representations of the matrix $R$ for different $\gamma_1, \gamma_2, \gamma_3$:
\begin{gather}
\label{threeR11}
R=\pm\omega (e_3+e_8) \\[1 ex]
\label{threeR12}
R=\varepsilon e_2+re_3\pm \omega e_8 \\[1 ex]
\label{threeR13}
R=\pm\omega (e_3-e_8) \\[1 ex]
\label{threeR14}
R=\frac12(r\pm\omega)e_0+\frac12(r\mp\omega)e_3+\varepsilon e_5-\frac12(3r\pm\omega)e_8 \\[1 ex]
\label{threeR15}
R=\pm\omega (e_0-2e_8) \\[1 ex]
\label{threeR16}
R=\frac12(r\pm\omega)e_0-\frac12(r\mp\omega)e_3+\varepsilon e_7-\frac12(3r\pm\omega)e_8
\end{gather}
and for $\gamma_1=\gamma_2\neq \gamma_3$:
\begin{gather}
\label{threeR21}
R=\pm\omega(e_1\pm e_8) \\[1 ex]
\label{threeR22}
R=\frac12(r\pm\omega)e_0+\frac12(r\mp\omega)e_3+\varepsilon e_5-\frac12(3r\pm\omega) e_8
\end{gather}
where $p, r, \phi, \varepsilon$ --- real constants, $p\neq0$ and $r^2+\varepsilon^2=\omega^2$.
In the (\ref{threeR21}) case, signs before $e_1$ and $e_8$ are chosen independently.

Below the list of nonequivalent representations of the matrix $P$ is presented.

For different $\gamma_1, \gamma_2, \gamma_3$, matrix $R$, that is defined by formula (\ref{threeR11}), corresponding to~$Q$,
that is defined by formulas (\ref{threeQ1})-(\ref{threeQ27}), (\ref{threeQ210})-(\ref{threeQ32})
\begin{gather}
\label{threePfirst}
P=\phi e_1\!\sqrt{\sec(\lambda x+\gamma_1)\sec(\lambda x+\gamma_2)}+pe_6\!\sqrt{\sec(\lambda x+\gamma_2)\sec(\lambda x+\gamma_3)} \\[1 ex]
P=\phi e_1\!\sqrt{\sech(\lambda x+\gamma_1)\sech(\lambda x+\gamma_2)}+pe_6\!\sqrt{\sech(\lambda x+\gamma_2)\sech(\lambda x+\gamma_3)} \\[1 ex]
P=\phi e_1\!\sqrt{\csch(\lambda x+\gamma_1)\csch(\lambda x+\gamma_2)}+pe_6\!\sqrt{\csch(\lambda x+\gamma_2)\csch(\lambda x+\gamma_3)} \\[1 ex]
P=\phi e_1\!\sqrt{\sech(\lambda x+\gamma_1)\sech(\lambda x+\gamma_2)}+pe_6\!\sqrt{\sech(\lambda x+\gamma_2)\csch(\lambda x+\gamma_3)} \\[1 ex]
P=\phi e_1\!\sqrt{\csch(\lambda x+\gamma_1)\csch(\lambda x+\gamma_2)}+pe_6\!\sqrt{\csch(\lambda x+\gamma_2)\sech(\lambda x+\gamma_3)} \\[1 ex]
P=\phi e_1\!\sqrt{\sech(\lambda x+\gamma_1)\sech(\lambda x+\gamma_2)}+pe_6\!\sqrt{\sech(\lambda x+\gamma_2)\exp(\pm\lambda x)} \\[1 ex]
P=\phi e_1\!\sqrt{\csch(\lambda x+\gamma_1)\csch(\lambda x+\gamma_2)}+pe_6\!\sqrt{\csch(\lambda x+\gamma_2)\exp(\pm\lambda x)} \\[1 ex]
P=\phi e_1\!\sqrt{\exp(\pm\lambda x)\sech(\lambda x+\gamma_2)}+pe_6\!\sqrt{\sech(\lambda x+\gamma_2)\csch(\lambda x+\gamma_3)} \\[1 ex]
P=\phi e_1+pe_6\!\sqrt{\sech(-\lambda x)\sech(\lambda x+\gamma_3)} \\[1 ex]
P=\phi e_1+pe_6\!\sqrt{\sech(-\lambda x)\csch(\lambda x+\gamma_3)} \\[1 ex]
P=\frac{\phi e_1}{\!\sqrt{(x+\gamma_1)(x+\gamma_2)}}+\frac{pe_6}{\!\sqrt{(x+\gamma_2)(x+\gamma_3)}} \\[1 ex]
P=\frac{\phi e_1}{\!\sqrt{(x+\gamma_1)(x+\gamma_2)}}+\frac{pe_6}{\!\sqrt{x+\gamma_2}}
\end{gather}

For different $\gamma_1, \gamma_2, \gamma_3$, matrix $R$, that are defined by formula (\ref{threeR12}), corresponding to~$Q$,
that is defined by formulas (\ref{threeQ1})-(\ref{threeQ27}), (\ref{threeQ210})-(\ref{threeQ32})
\begin{gather}
P=pe_1\!\sqrt{\sec(\lambda x+\gamma_1)\sec(\lambda x+\gamma_2)} \\[1 ex]
P=pe_1\!\sqrt{\sech(\lambda x+\gamma_1)\sech(\lambda x+\gamma_2)} \\[1 ex]
P=pe_1\!\sqrt{\csch(\lambda x+\gamma_1)\csch(\lambda x+\gamma_2)} \\[1 ex]
P=pe_1\!\sqrt{\sech(\lambda x+\gamma_1)\sech(\lambda x+\gamma_2)} \\[1 ex]
P=pe_1\!\sqrt{\csch(\lambda x+\gamma_1)\csch(\lambda x+\gamma_2)} \\[1 ex]
P=pe_1\!\sqrt{\sech(\lambda x+\gamma_1)\sech(\lambda x+\gamma_2)} \\[1 ex]
P=pe_1\!\sqrt{\csch(\lambda x+\gamma_1)\csch(\lambda x+\gamma_2)} \\[1 ex]
P=pe_1\!\sqrt{\exp(\pm\lambda x)\sech(\lambda x+\gamma_2)} \\[1 ex]
P=pe_1 \\[1 ex]
P=pe_1 \\[1 ex]
P=\frac{pe_1}{\!\sqrt{(x+\gamma_1)(x+\gamma_2)}} \\[1 ex]
P=\frac{pe_1}{\!\sqrt{(x+\gamma_1)(x+\gamma_2)}}
\end{gather}

For different $\gamma_1, \gamma_2, \gamma_3$, matrix $R$, that is defined by formula (\ref{threeR13}), corresponding to~$Q$,
that is defined by formulas (\ref{threeQ27}), (\ref{threeQ210}), (\ref{threeQ211})
\begin{gather}
P=\phi e_1\!\sqrt{\exp(\pm\lambda x)\sech(\lambda x+\gamma_2)}+pe_4\!\sqrt{\exp(\pm\lambda x)\csch(\lambda x+\gamma_3)} \\[1 ex]
P=\phi e_1+pe_4\!\sqrt{\exp(\lambda x)\sech(\lambda x+\gamma_3)} \\[1 ex]
P=\phi e_1+pe_4\!\sqrt{\exp(\lambda x)\csch(\lambda x+\gamma_3)}
\end{gather}

For different $\gamma_1, \gamma_2, \gamma_3$, matrix $R$, that is defined by formula (\ref{threeR14}), corresponding to~$Q$,
that is defined by formulas (\ref{threeQ23})-(\ref{threeQ27}), (\ref{threeQ210}), (\ref{threeQ211}), (\ref{threeQ32})
\begin{gather}
P=pe_4\!\sqrt{\sech(\lambda x+\gamma_1)\csch(\lambda x+\gamma_3)} \\[1 ex]
P=pe_4\!\sqrt{\csch(\lambda x+\gamma_1)\sech(\lambda x+\gamma_3)} \\[1 ex]
P=pe_4\!\sqrt{\sech(\lambda x+\gamma_1)\exp(\pm\lambda x)} \\[1 ex]
P=pe_4\!\sqrt{\csch(\lambda x+\gamma_1)\exp(\lambda x)} \\[1 ex]
P=pe_4\!\sqrt{\exp(\pm\lambda x)\csch(\lambda x+\gamma_3)} \\[1 ex]
P=pe_4\!\sqrt{\exp(\lambda x)\sech(\lambda x+\gamma_3)} \\[1 ex]
P=pe_4\!\sqrt{\exp(\lambda x)\csch(\lambda x+\gamma_3)} \\[1 ex]
P=\frac{pe_4}{\!\sqrt{x+\gamma_1}}
\end{gather}

For different $\gamma_1, \gamma_2, \gamma_3$, matrix $R$, that is defined by formula (\ref{threeR15}), corresponding to~$Q$,
that is defined by formulas (\ref{threeQ23})-(\ref{threeQ27}), (\ref{threeQ210}), (\ref{threeQ211}), (\ref{threeQ32})
\begin{gather}
P=\phi e_4\!\sqrt{\sech(\lambda x+\gamma_1)\csch(\lambda x+\gamma_3)}+pe_6\!\sqrt{\sech(\lambda x+\gamma_2)\csch(\lambda x+\gamma_3)} \\[1 ex]
P=\phi e_4\!\sqrt{\csch(\lambda x+\gamma_1)\sech(\lambda x+\gamma_3)}+pe_6\!\sqrt{\csch(\lambda x+\gamma_2)\sech(\lambda x+\gamma_3)} \\[1 ex]
P=\phi e_4\!\sqrt{\sech(\lambda x+\gamma_1)\exp(\pm\lambda x)}+pe_6\!\sqrt{\sech(\lambda x+\gamma_2)\exp(\pm\lambda x)} \\[1 ex]
P=\phi e_4\!\sqrt{\csch(\lambda x+\gamma_1)\exp(\lambda x)}+pe_6\!\sqrt{\csch(\lambda x+\gamma_2)\exp(\pm\lambda x)} \\[1 ex]
P=\phi e_4\!\sqrt{\exp(\pm\lambda x)\csch(\lambda x+\gamma_3)}+pe_6\!\sqrt{\sech(\lambda x+\gamma_2)\csch(\lambda x+\gamma_3)} \\[1 ex]
P=\phi e_4\!\sqrt{\exp(\lambda x)\sech(\lambda x+\gamma_3)}+pe_6\!\sqrt{\exp(-\lambda x)\sech(\lambda x+\gamma_3)} \\[1 ex]
P=\phi e_4\!\sqrt{\exp(\lambda x)\csch(\lambda x+\gamma_3)}+pe_6\!\sqrt{\exp(-\lambda x)\csch(\lambda x+\gamma_3)} \\[1 ex]
P=\frac{\phi e_4}{\!\sqrt{x+\gamma_1}}+\frac{pe_6}{\!\sqrt{x+\gamma_2}}
\end{gather}

For different $\gamma_1, \gamma_2, \gamma_3$, matrix $R$, that is defined by formula (\ref{threeR16}), corresponding to~$Q$,
that is defined by formulas (\ref{threeQ27}), (\ref{threeQ210}), (\ref{threeQ211})
\begin{gather}
P=pe_6\!\sqrt{\exp(\pm\lambda x)\csch(\lambda x+\gamma_3)} \\[1 ex]
P=pe_6\!\sqrt{\exp(\lambda x)\sech(\lambda x+\gamma_3)} \\[1 ex]
P=pe_6\!\sqrt{\exp(\lambda x)\csch(\lambda x+\gamma_3)}
\end{gather}

For $\gamma_1=\gamma_2\neq \gamma_3$, matrix $R$, that is defined by formula (\ref{threeR21}), corresponding to~$Q$,
that is defined by formulas (\ref{threeQ1})-(\ref{threeQ26}), (\ref{threeQ28}), (\ref{threeQ29}), (\ref{threeQ31})-(\ref{threeQ33}),
where the sign in the matrix $P$ before $e_6$ --- opposite to the chosen sign in the matrix $R$ before $e_8$
\begin{gather}
P=pe_3\sec(\lambda x+\gamma_1)+\phi(e_4\mp e_6)\!\sqrt{\sec(\lambda x+\gamma_1)\sec(\lambda x+\gamma_3)} \\[1 ex]
P=pe_3\sech(\lambda x+\gamma_1)+\phi(e_4\mp e_6)\!\sqrt{\sech(\lambda x+\gamma_1)\sech(\lambda x+\gamma_3)} \\[1 ex]
P=pe_3\csch(\lambda x+\gamma_1)+\phi(e_4\mp e_6)\!\sqrt{\csch(\lambda x+\gamma_1)\csch(\lambda x+\gamma_3)} \\[1 ex]
P=pe_3\sech(\lambda x+\gamma_1)+\phi(e_4\mp e_6)\!\sqrt{\sech(\lambda x+\gamma_1)\csch(\lambda x+\gamma_3)} \\[1 ex]
P=pe_3\csch(\lambda x+\gamma_1)+\phi(e_4\mp e_6)\!\sqrt{\csch(\lambda x+\gamma_1)\sech(\lambda x+\gamma_3)} \\[1 ex]
P=pe_3\sech(\lambda x+\gamma_1)+\phi(e_4\mp e_6)\!\sqrt{\sech(\lambda x+\gamma_1)\exp(\pm\lambda x)} \\[1 ex]
P=pe_3\csch(\lambda x+\gamma_1)+\phi(e_4\mp e_6)\!\sqrt{\csch(\lambda x+\gamma_1)\exp(\pm\lambda x)} \\[1 ex]
P=pe_3\exp(\pm\lambda x)+\phi(e_4\mp e_6)\!\sqrt{\exp(\pm\lambda x)\sech(\lambda x+\gamma_3)} \\[1 ex]
P=pe_3\exp(\pm\lambda x)+\phi(e_4\mp e_6)\!\sqrt{\exp(\pm\lambda x)\csch(\lambda x+\gamma_3)} \\[1 ex]
P=\frac{pe_3}{x+\gamma_1}+\frac{\phi(e_4\mp e_6)}{\!\sqrt{(x+\gamma_1)(x+\gamma_3)}} \\[1 ex]
P=\frac{pe_3}{x+\gamma_1}+\frac{\phi(e_4\mp e_6)}{\!\sqrt{x+\gamma_1}} \\[1 ex]
P=pe_3+\frac{\phi(e_4\mp e_6)}{\!\sqrt{x+\gamma_3}}
\end{gather}

For $\gamma_1=\gamma_2\neq \gamma_3$, matrix $R$, that is defined by formula (\ref{threeR22}), corresponding to
$Q$, that is defined by formulas (\ref{threeQ1})-(\ref{threeQ26}), (\ref{threeQ28}), (\ref{threeQ29}), (\ref{threeQ31})-(\ref{threeQ33})
\begin{gather}
P=pe_4\!\sqrt{\sec(\lambda x+\gamma_1)\sec(\lambda x+\gamma_3)} \\[1 ex]
P=pe_4\!\sqrt{\sech(\lambda x+\gamma_1)\sech(\lambda x+\gamma_3)} \\[1 ex]
P=pe_4\!\sqrt{\csch(\lambda x+\gamma_1)\csch(\lambda x+\gamma_3)} \\[1 ex]
P=pe_4\!\sqrt{\sech(\lambda x+\gamma_1)\csch(\lambda x+\gamma_3)} \\[1 ex]
P=pe_4\!\sqrt{\csch(\lambda x+\gamma_1)\sech(\lambda x+\gamma_3)} \\[1 ex]
P=pe_4\!\sqrt{\sech(\lambda x+\gamma_1)\exp(\pm\lambda x)} \\[1 ex]
P=pe_4\!\sqrt{\csch(\lambda x+\gamma_1)\exp(\pm\lambda x)} \\[1 ex]
P=pe_4\!\sqrt{\exp(\pm\lambda x)\sech(\lambda x+\gamma_3)} \\[1 ex]
P=pe_4\!\sqrt{\exp(\pm\lambda x)\csch(\lambda x+\gamma_3)} \\[1 ex]
P=\frac{pe_4}{\!\sqrt{(x+\gamma_1)(x+\gamma_3)}} \\[1 ex]
P=\frac{pe_4}{\!\sqrt{x+\gamma_1}} \\[1 ex]
\label{threePlast}
P=\frac{pe_4}{\!\sqrt{x+\gamma_3}}
\end{gather}

The corresponding three-dimensional matrix superpotentials can be easily recreated by means of the formula
(\ref{superpot}).

\section{Conclusions \label{concl}}

The article intended to find three-dimensional matrix superpotentials, that correspond to
exactly integrable systems of Schr\"odinger equations. In the general form the problem
remains rather complex to analyze, but it was solved by limiting the class of the considered
superpotentials by a simple case~(\ref{superpot}).

The list of considered three-dimensional superpotentials is presented by means of matrices
$Q$ (formulas (\ref{threeQ1})-(\ref{threeQ33})), $R$ (formulas (\ref{threeR11})-(\ref{threeR22}))
and $P$ (formulas (\ref{threePfirst})-(\ref{threePlast})).
Corresponding superpotentials is easy to get explicitly by means of formula~(\ref{superpot})
beginning with formulas for the matrix $P$.

The received results can be interesting for physicists, in order to find new exactly integrable physical models.

Even though the problem was solved entirely, the achieved results are limited by the form of superpotential (\ref{superpot})
and the dimension of the superpotential. In the upcoming works, a more general class of superpotentials will be considered.

\subsection*{Acknowledgement}
The author thanks to Prof. Anatoly Nikitin for useful discussions and valuable comments.

\end{document}